\documentclass[aps,prc,preprint,superscriptaddress,showpacs]{revtex4}

\usepackage{graphicx}
\usepackage{dcolumn}
\usepackage{bm}
\usepackage{amsmath}

\begin{document}
\begin{center}
{\large\bf Thomas-Fermi approximation to pairing in finite Fermi systems. 
The weak coupling regime}

\centerline{X. Vi\~nas$^{1}$, P. Schuck$^{2,3}$, and M. Farine$^{4}$}

%\address{
{\it $^{1}$Departament d'Estructura i Constituents de la Mat\`eria
and Institut de Ci\`encies del Cosmos, Facultat de F\'{\i}sica,
Universitat de Barcelona, Diagonal {\sl 647}, {\sl E-08028} Barcelona,
Spain} \\
{\it $^{2}$Institut de Physique Nucl\'eaire, IN2P3-CNRS, Universit\'e Paris-Sud,
\\F-91406 Orsay-C\'edex, France} \\
{\it $^{3}$ Laboratoire de Physique et Mod\'elisation des Milieux Condens\'es,
CNRS and Universit\'e Joseph Fourier, 25 Avenue des Martyrs, Bo\^{i}te Postale 166,
F-38042 Grenoble Cedex 9, France}\\
{\it $^{4}$ Ecole des Mines de Nantes, Universit\'e Nantes,
4, rue Alfred Kastler B.P. 20722\\
{\sl 44307}  Nantes-C\'edex 3, France}
%}
\end{center}
%\ead{xavier@ecm.ub.es}

\begin{abstract}
We present a new semiclassical theory for describing pairing in
finite Fermi systems. It is based on taking the $\hbar \to 0$, i.e.
Thomas-Fermi, limit of the gap equation written in the basis of the
mean field (weak coupling). In addition to the position dependence
of the Fermi momentum, the size dependence of the pairing force is
also taken into account in this theory. Along isotopic chains the 
Thomas-Fermi gaps average the well known arch structure shown by 
the quantal gaps. This structure can be almost recovered in our 
formalism if some shell fluctuations are included in the level density. 
We point out that at the drip line nuclear pairing is strongly reduced.
This fact is illustrated with the behaviour of the gap in the inner crust 
of neutron stars.

\end{abstract}

\section{Introduction}

Semiclassical approaches to finite Fermi systems provide a
very efficient way of extracting the average behaviour of relevant 
physical quantities which characterize such  systems. 
The most well known example is the celebrated Droplet Model and 
its extensions developed by Myers and Swiatecki which 
describe nicely the average behaviour of the nuclear masses \cite{MS}. 
Semiclassical techniques have also been applied to study
the average behaviour of other properties such as inertias \cite{far00}, 
charge radii \cite{Piek10}, one- and two-body matrix elements\cite{vin03}, 
etc. Our aim here is to present a new Thomas-Fermi (TF) theory, i.e. 
the $\hbar \to 0$ limit, for describing the average trends of 
the effect of pairing correlations in finite Fermi systems. 
Semiclassical approaches to the pairing problem can be
of interest in scenarios like, e.g., cold atomic gases
where the huge number of particles makes the full quantal calculation 
numerically very complicated. Also these calculations are useful if 
one is only interested in the average behaviour of the pairing gap, 
as is the case of the pairing term in the nuclear mass formula.
The Local Density Approximation (LDA) is the standard semiclassical technique
for dealing with the average behaviour of the pairing which was developed by 
Schuck and collaborators more than twenty years ago \cite{kuch89}.  
In LDA one considers the BCS equations in infinite homogeneuous matter and 
replaces the Fermi momentum $k_F$ by its local version in terms of the density.
The validity of LDA applied to the pairing problem is restricted on one hand
to situations where the local Fermi wavelength $2 \pi/k_F({\bf R})$ is small 
as compared with the distance where the 
mean field potential varies appreciably. In the case of a harmonic oscillator
potential  
$V({\bf R})=m \omega^2 R^2/2$ this distance is the so-called  
oscillator length defined as $l=\sqrt{\hbar/m \omega}$. On the other hand, a 
second length scale 
introduced by pairing is the coherence length $\xi$ which measures 
the extension 
of the Cooper pairs. The validity of LDA in the pairing case also implies that 
the coherence length be smaller than the oscillator legth, 
i.e. $\xi/l < 1$, which
is usually equivalent to the condition $\Delta/\hbar \omega > 1$
, where $\Delta$ is the gap in the single-particle spectrum. 
The condition $\xi/l < 1$ is always violated in the outer tail of the
surface because the LDA coherence length behaves as $\xi \sim \Delta^{-1}$ and
the gap vanishes in this region. In spite of these deficiences, 
integrated quantities as pairing energies may be quite accurate when 
considered on average \cite{kuch89}.

In this contribution we present a novel TF theory for pairing which 
improves the LDA. This theory can be applied in the weak pairing 
regime where the chemical potential $\mu$ and the Fermi energy
$\varepsilon_F$ have  similar values and  
$\Delta/\mu << 1$. This TF theory works, for the average, in 
the region $\Delta < \hbar \omega$, where LDA  generally  fails.

An important point that will be discussed in this contribution is
the possible quenching of the pairing gap when approaching the
drip line. This fact is well documented in the recent nuclear physics
literature \cite{ham05,san04,bal07,gra08,cha10}
and we will see, using some examples, 
that this effect is a general feature when superfluid (superconducting) 
fermions come in a finite confining potential to an overflow situation.   

The contribution is organized as follows. The basic theory is 
presented in the second section. The main results are discussed in the 
third section. Our conclusions are laid out in the last section.
  
\section{Basic theory}

It is well known that in the Hartree-Fock-Bogoliubov theory single-particle
density matrix and the pairing tensor or anomalous density matrix 
are simultaneously 
diagonalized by the so-called canonical or natural basis, 
$|n_c \rangle$ \cite{RS}.
As far as in this work we are only interested in the weak coupling limit
where the gap is small as compared with the Fermi energy, i.e. $\Delta/\mu < 1$,
one can replace with only small error the canonical basis by the basis of the 
normal, non-superfluid mean field (HF) Hamiltonian. In this situation the
gap equation reduces to its BCS approximation and can be written as
%$\hat \kappa$:
%\begin{equation}
%\hat \rho|n_c \rangle = v_n^2|n_c \rangle, \quad 
%\hat \kappa|n_c \rangle = u_nv_n|n_c \rangle,
%\label{eq1}
%\end{equation}
%where the eigenvalues $v_n^2$ have the meaning of occupation numbers and the 
%amplitudes $u_n, v_n$, normalised as $u_n^2 + v_n^2 = 1$,  are analogous to 
%the ones 
%also used in BCS theory \cite{RS}. 
\begin{equation}
\Delta_{n} =- \sum_{n'}V_{nn'}\frac{\Delta_{n'}}{2E_{n'}},
\label{eq2}
\end{equation}
where $V_{nn'} = \langle  n {\bar n}|v|n' {\bar n'} \rangle$   
is the matrix element of the interaction with $|{\bar n}\rangle$ the
time reversed state of $|n\rangle$ and
$E_{n} = [(\epsilon_{n} - \mu)^2 + \Delta_{n}^2]^{1/2}$ the quasi-particle 
energies, with $\epsilon_{n}$
the diagonal elements of the normal mean field Hamiltonian \cite{RS} written in 
the basis of the standard mean field hamiltonian, that is  
$H|n \rangle = \epsilon_n|n \rangle$.

At equilibrium and for time reversal invariant systems canonical conjugation 
and time reversal operation are related by
%\begin{equation}
$\langle{\bf r}|{\bar n}\rangle =
\langle n|{\bf r}\rangle \Rightarrow
\langle{\bf r}_1 {\bf r}_2|n {\bar n}\rangle =
\langle{\bf r}_1|{\hat \rho_n}|{\bf r}_2\rangle$,
%\label{eq2a}
%\end{equation}
where $\hat \rho_n = |n\rangle \langle n|$ is the density matrix corresponding
to the state $|n\rangle$. Therefore the pairing matrix element can be writeen as:
\begin{equation}
V_{n n'} = \langle  n {\bar n}|v|n' {\bar n'} \rangle
=\int d {\bf r_1} d {\bf r_2} d{\bf r'_1} d {\bf r'_2} \langle{\bf r'}_1|{\hat \rho_n}|{\bf r}_1 \rangle
\langle{\bf {r_1}} {\bf {r_2}} \vert v \vert {\bf {r_1'}} {\bf {r_2'}}\rangle
\langle{\bf r'}_2|{\hat \rho_n}|{\bf r}_2 \rangle.
\label{eq2b}
\end{equation}
The density matrix $\hat \rho_n$ fulfills the Sch\"odinger equation
\begin{equation}
(H - \epsilon_n)\hat \rho_n = 0,
\label{eq3}
\end{equation}
therefore we can write $\Delta_n = Tr[\hat \Delta \hat \rho_n]$ and
$\epsilon_n = Tr[H \hat \rho_n]$ and consequently
the state dependence of the gap equation (\ref{eq2}) is
fully expressed through the density matrix $\hat \rho_n$.

Performing the Wigner transform (WT) of Eq.(\ref{eq3}) and taking into 
account that the WT of the product of two single-particle 
operators ${\hat A}$ and ${\hat B}$ equals, to lowest order in 
$\hbar$, the c-number product of the corresponding WT's, i.e. 
$A({\bf R},{\bf p})B({\bf R},{\bf p})$, one easily obtains  
the $\hbar \to 0$ limit of Eq.(\ref{eq3}) \cite{RS}
\begin{equation}
(H_{cl.} - \epsilon)f_{\epsilon}({\bf R},{\bf p}) =0,
\label{eq5}
\end{equation}
where $H_{cl.} = \frac{p^2}{2m^*({\bf R})} + V({\bf R})$ is the classical
Hamiltonian which contains a local
mean field potential $V({\bf R})$ and a position dependent
effective mass $m^*({\bf R})$ and $f_{\epsilon}({\bf R},{\bf p})$ is the Wigner 
transform of $\hat \rho_n$.
Equation (\ref{eq5}) has to be read in the sense of distributions. 
Taking into account that $x\delta(x)=0$ one obtains the normalized distribution 
function
\begin{equation}
f_E({\bf R},{\bf p}) = \frac{1}{g^{TF}(E)}\delta(E - H_{cl.}) + O(\hbar^2),
\label{eq6}
\end{equation}
which corresponds to the Thomas-Fermi (TF) approximation of the normalized 
on-shell or spectral density matrix \cite{vin03}. Its norm is equal 
to the level density 
(without spin-isospin degeneracy):
\begin{equation}
g^{TF}(E) = \frac{1}{(2\pi \hbar)^3} \int  d {\bf R} d {\bf p} \delta(E -
H_{cl.}).
\label{eq7}
\end{equation}

The semiclassical pairing matrix element can then be written as 
\cite{vin03}:
\begin{equation}
V(E,E') =
\int \frac{d {\bf R} d{\bf p}}{(2 \pi \hbar)^3}
\int \frac{d{\bf R'} d{\bf p'}}{(2 \pi \hbar)^3}
f_E({\bf R},{\bf p}) f_{E'}({\bf R'},{\bf p'})
v({\bf R},{\bf p};{\bf R'},{\bf p'}),
\label{eq10}
\end{equation}
where $v({\bf R},{\bf p};{\bf R'},{\bf p'})$ is the double WT of
$<{\bf {r_1}} {\bf {r_2}} \vert v \vert {\bf {r_1'}} {\bf {r_2'}}>$. 
For a local translationally invariant force this matrix element reduces to
$v({\bf R},{\bf p};{\bf R'},{\bf p'})=\delta({\bf R}-{\bf R'})
v({\bf p}-{\bf p'})$ with $v({\bf p}-{\bf p'})$ the Fourier transform of the force
$v({\bf r}-{\bf r'})$ in coordinate space. 
%In the particular case of 
%a density dependent zero range force, one obtains 
%$v_0(\rho({\bf R}))\delta({\bf R}-{\bf R}')$, 
%whith $v_0(\rho)$ the density dependent pairing strength \cite{gar99}.

The gap equation in the TF approximation is obtained by replacing in (\ref{eq2})
$\hat \rho_n$ and $V_{nn'}$ by their corresponding  semiclassical counterparts 
Eqs. (\ref{eq6}) and (\ref{eq10}) respectively. In this way the TF gap equation 
reads
\begin{equation}
\Delta(E) = \int_0^{\infty} dE' g^{TF}(E') V(E,E')\frac{\Delta(E')}
{2\sqrt{(E'-\mu)^2 + \Delta^2(E')}} ,
\label{eq8}
\end{equation}
Eqs.(\ref{eq8})-(\ref{eq10}) can  readily be solved for a given mean field
and the chemical potential is fixed by the usual particle number condition.

\section{Results}

As a realistic application of our TF theory  we analyze the semiclassical 
pairing gaps as a function of mass 
number along the tin isotopic chain from $^{100}$Sn to $^{132}$Sn.   
To this end we use the D1S Gogny force \cite{D1S} for both, mean field and pairing 
fields. The main ingredients for solving the semiclassical pairing equation 
(\ref{eq8}) are the on-shell density matrix $f_E({\bf R},{\bf p})$ 
(\ref{eq6}), which depends on the classical Hamiltonian $H_{cl}$ that is 
determined by the effective mass $m^*({\bf R})$  and the mean 
field $V({\bf R})$.  
These two quantities, namely $m^*({\bf R})$ and $V({\bf R})$,
are obtained through the Extended Thomas-Fermi (ETF) theory
for finite-range non-relativistic interactions \cite{cent98,soub00}.
%We compute $m^*({\bf R})$ and $V({\bf R})$ self-consistently and the 
%corresponding results are displayed by dashed lines 
%in the left panel of Figure 2 in the case 
%of the nucleus $^{116}$Sn. 
Using these quantities as input, one obtains 
the level density (\ref{eq7}) and the pairing matrix element (\ref{eq10}) 
which allow to solve the gap equation (\ref{eq8}) in our TF approximation. 
%As we have shown before, these TF gaps will be very smooth as a function of 
%particle number and, consequently, the quantal arch structure will be washed 
%out.
%Of course, the TF solution of the gap only can yield a monotoneous 
%behavior as a function of $\mu$.
%We can try to recover the quantal. i.e. arch structure over the shell 
%by introducing additional quantal
%information. As it is known for pairing, quantal fluctuations are  
%more important in the level density than in the pairing matrix elements. 
%To do that we proceed as follows. As it has been explained 
%in Refs.\cite{soub00,soub03}, 
As explained in Refs. \cite{soub00,soub03}, the ETF energy density functional 
can be transformed, inspired by the Kohn-Sham scheme, into a quantal functional
from where the quantal average gaps are obtained. It should be 
noted that within this approximation the quantal functional associated to 
a finite-
range effective interaction becomes local \cite{soub00,soub03}. 
%The quantal $V({\bf R})$ and 
%$m^*({\bf R})$  obtained in this way are also displayed by solid lines in 
%Figure 2. We see, as expected, that the quantal oscillations are nicley 
%averaged by the ETF solutions. 
The quantal pairing gaps averaged with $u^2 v^2$ are depicted 
by circles in the left panel of Fig. 1 and show the typical arch structure. 
In the same panel we also display the semiclassical TF gap at the Fermi energy
by a thick solid line. We see that in this case the quantal arch structure 
completely 
disappears, as expected due the absence of shell effects in this case, and 
that the TF gaps decrease smoothly when the neutron number increases.
As it has been discussed in Ref. \cite{vin11}, quantal effects, i.e. 
the arch structure over the shell, can be almost recovered by introducing 
some additional quantal fluctuations in the level density and retaining 
the TF pairing matrix elements
(\ref{eq10}) in the gap equation (\ref{eq8}).
%we proceed as f$ollows.
%For a given nucleus, once the single-particle energy levels have been
%obtained, we build a fluctuating level density by taking a sum of
%Gaussians each one centered at a single-particle energy, with a
%width $\sigma=0.5$ MeV and with a strength such that the area below the
%Gaussian equals the degeneracy of each energy level (spherical symmetry
%is assumed). 
%In this way the fluctuating level density reads:
%\begin{equation}
%${\tilde g}(E) = \sum_{i=1}^{n_{tot}} g_{0,i}
%e^{-(\frac{E - \varepsilon_i}{\sigma})^2}$
%\end{equation}
%where $n_{tot}$ is the total number of single-particle energy levels
%considered. 
%This fluctuating level density ${\tilde g}(E)$ corresponding to the nucleus
%$^{116}$Sn is displayed in right panel of Figure 2, where we also show, for
%comparison, the smooth TF level density $g^{TF}(E)$.
% for the same nucleus.
%We solve the gap equation (\ref{eq8}) using this fluctuating level density 
%${\tilde g}(E)$ 
%but retaining the semiclassical matrix elements. 
The average gaps obtained in these conditions  
are displayed by diamonds in the left panel of Figure 2. We see that
the arch structure for the tin isotopic chain is recovered 
and that the quantal gaps are predicted quite well in this way.
%The quantal features of pairing are essentially recovered in this way.
\begin{figure}
\begin{center}
\includegraphics[height=7.5cm,angle=-90]{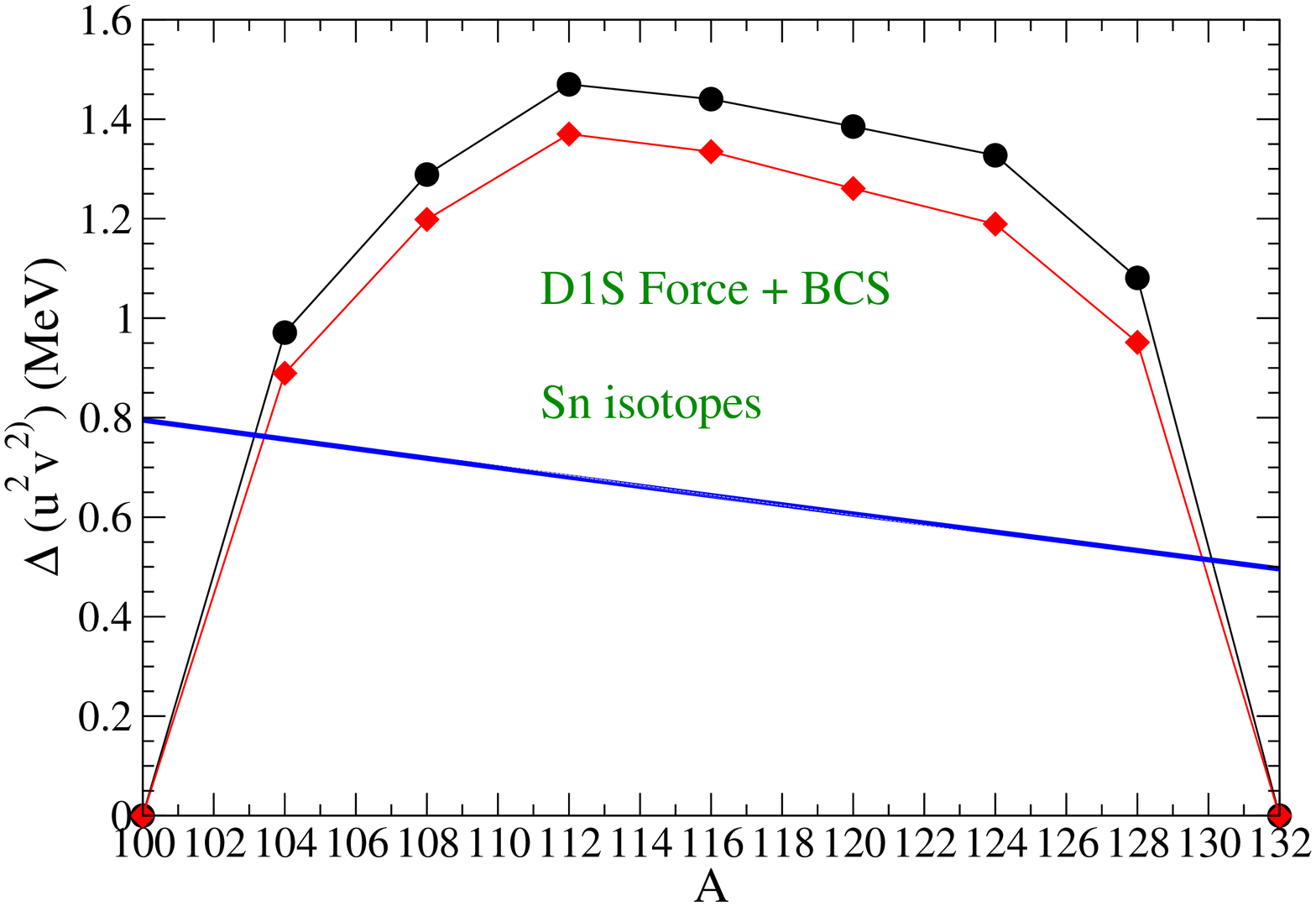}
\includegraphics[height=7.5cm,angle=-90]{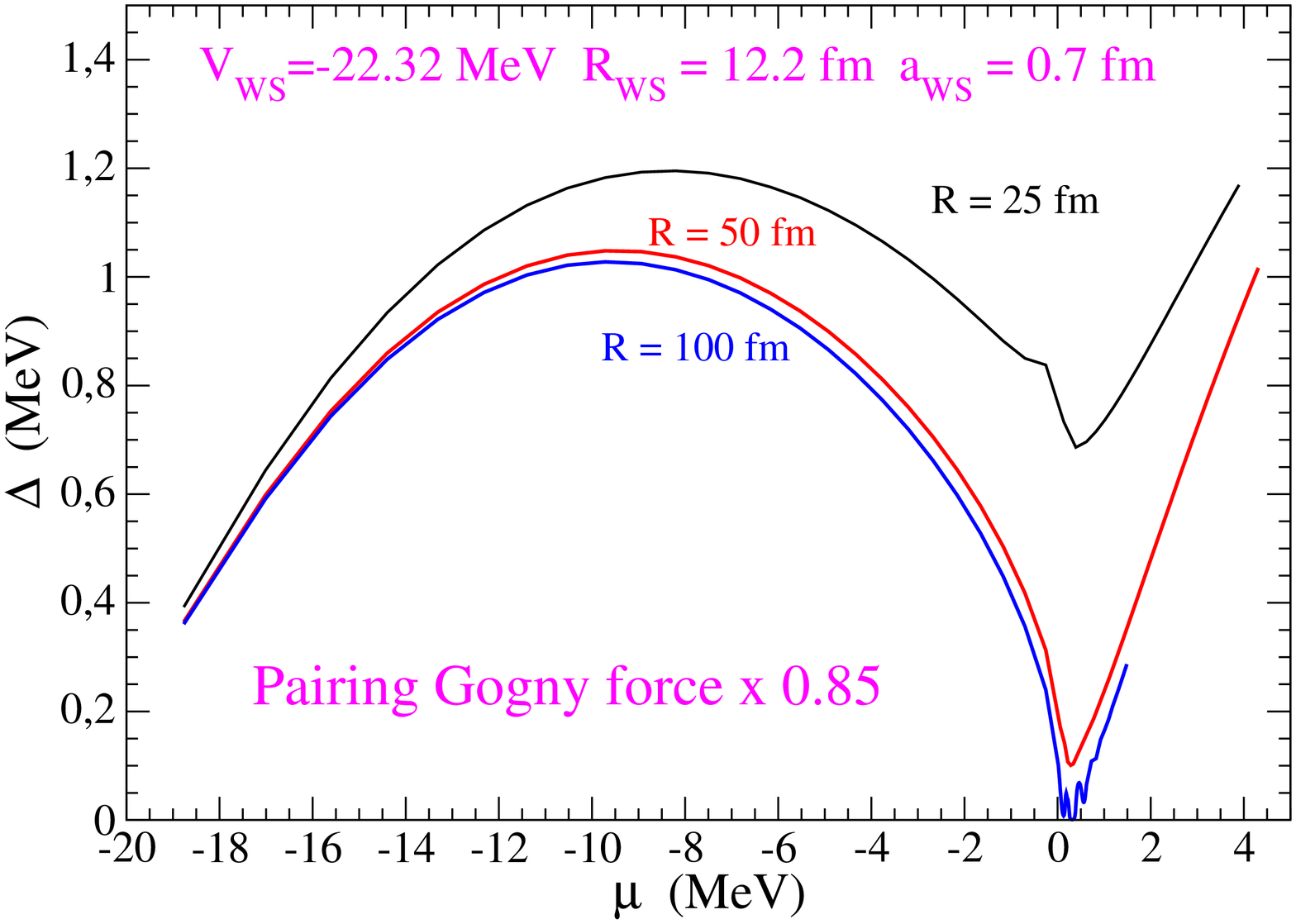}
\end{center}
\caption{\label{Fig1} Left: Average pairing gap along the Sn 
isotopic chain. See text for details. Right: Average TF gaps at the 
Fermi energy as a function of the chemical potential in a WS potential
computed in a box of radius $R$=25, 50, and 100 fm.}
 \end{figure}
\begin{figure}
\begin{center}
\includegraphics[height=7.5cm,angle=-90]{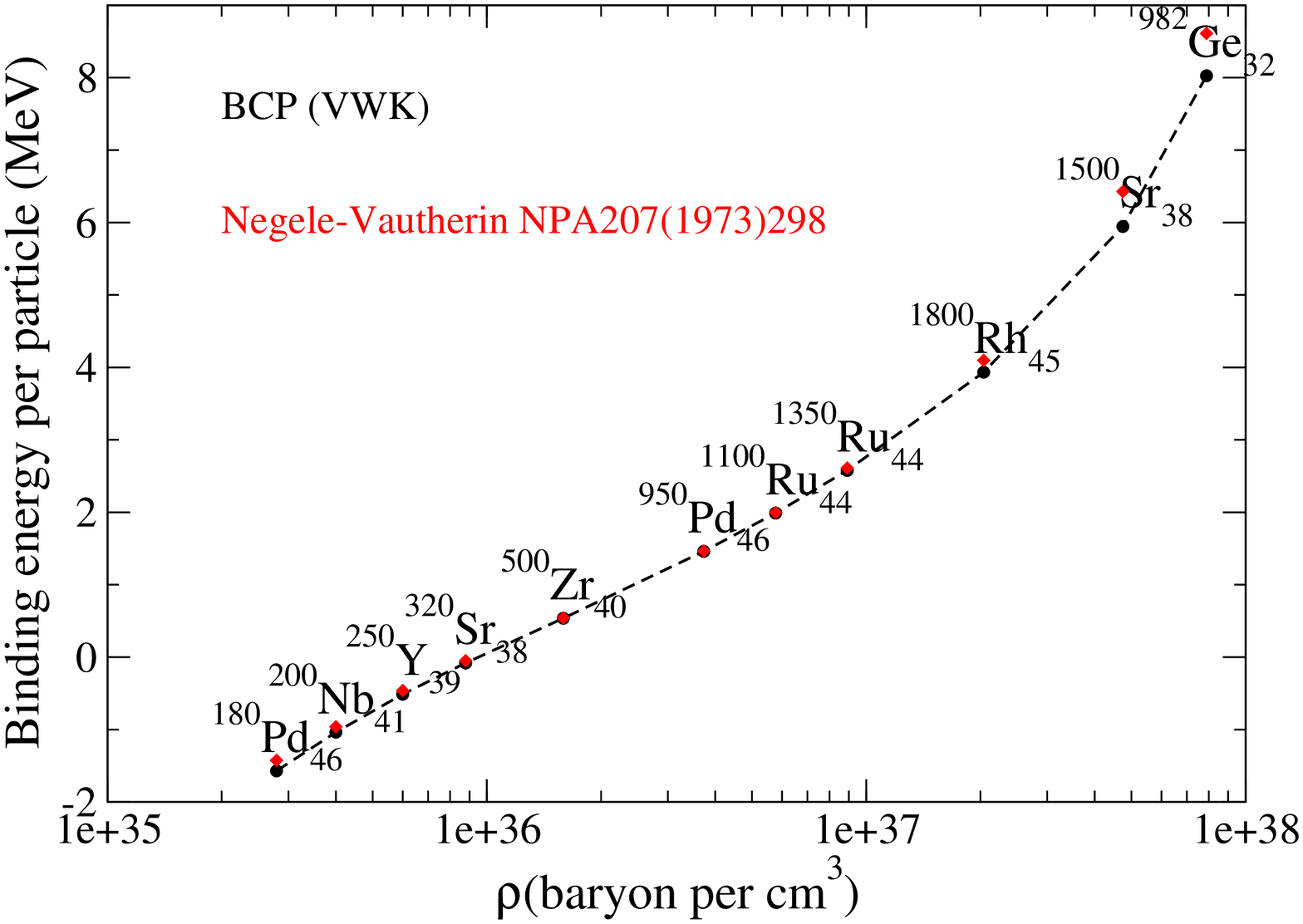}
\includegraphics[height=7.5cm,angle=-90]{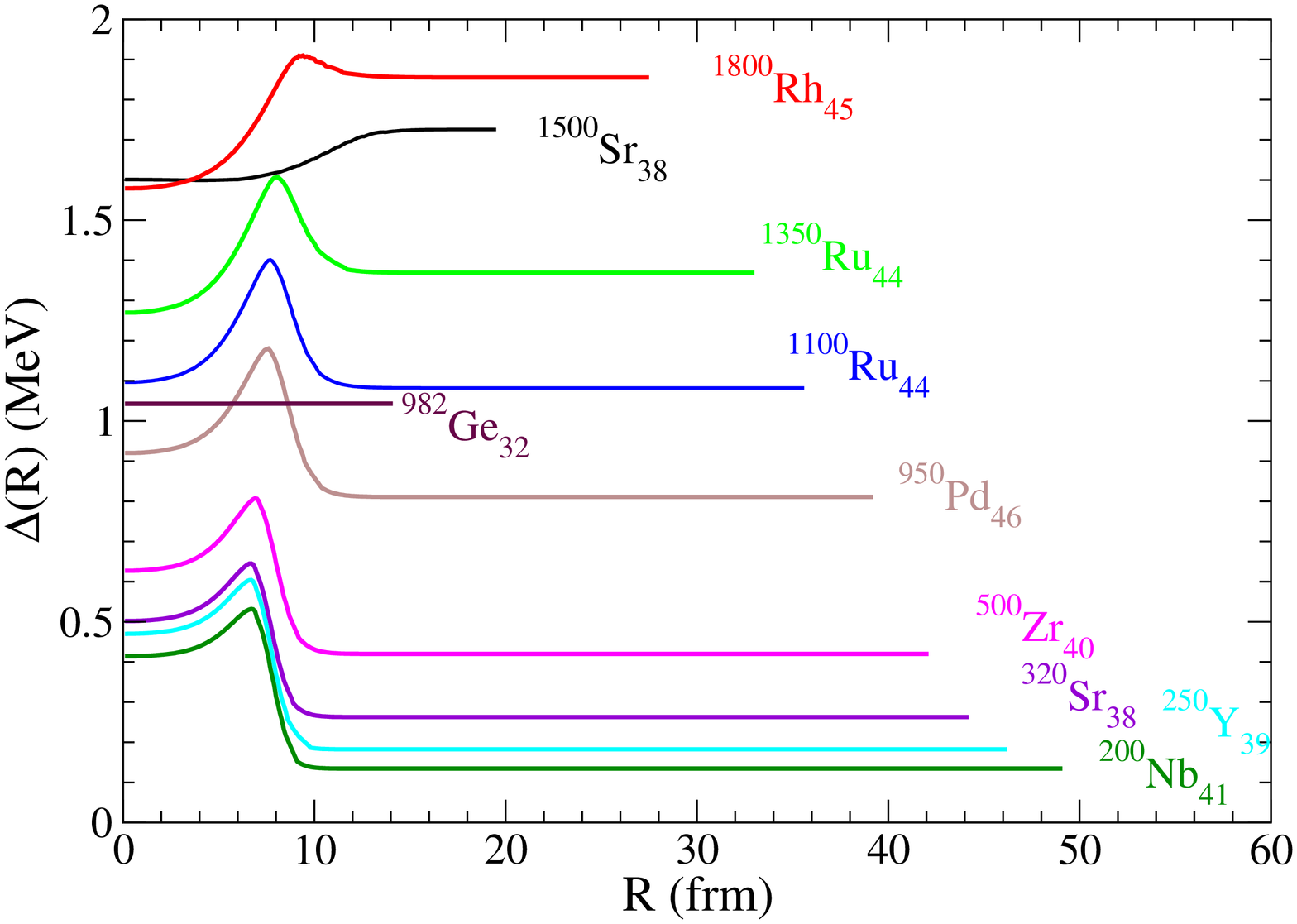}
\caption{\label{Fig2} Left: Energy per baryon as a
function of the average density in different WS cells corresponding to
the inner crust of neutron stars. Right: Radial dependence of TF
gap  in the analyzed WS cells. End points indicate radius of WS cells.}
\end{center}
\end{figure}
%{\bf We display in Figure 3} 
%The average quantal gaps  
%weighted with $u^2 v^2$ (circles) and the ones
%obtained with our TF theory for pairing using the fluctuating
%level density (diamonds) for the tin isotopic chain 
%as a function of the mass number are displayed in Figure 3.
%We see that by introducing ${\tilde g}(E)$ the
%quantal arch structure is recovered and that the semiclassical
%gaps obtained in this way reproduce quite accurately the quantal values.
%In the same Figure we also display the semiclassical averages of the gap
%computed with the smooth TF level density (solid line). We see that
%in this case the quantal arch structure is completely washed out  
%and that the semiclassical average gaps show a downward tendency with 
%increasing neutron number. \\

An important feature of the average TF gaps is that they show a downward trend 
with increasing neutron number along a given isotopic chain 
as it can be appreciated 
in the left panel of Fig. 1. To investigate the behaviour of the TF gap at 
the Fermi energy approaching the drip line, we show in the right panel of 
Fig. 1 the TF gap as a function of the chemical potential employing 
a Woods-Saxon potential as mean field \cite{shl92}. From this panel
it can be seen that in the TF limit the gap vanishes just at the drip line when 
the chemical potential equals the depth of the single-particle potential. 
Actually we put the Woods Saxon potential in a large external container 
with radius $R$ what can simulate a Wigner-Seitz (WS) cell as relevant in the 
inner crust of neutron stars, see below. We see that the gap becomes more 
and more suppressed at the edge of the Woods-Saxon potential 
for increasing values of the cell radius. Continuing filling the cell 
with a neutron gas, the gaps raise again. 
This quenching of the gap when approaching the drip line has also been studied
by Hamamoto \cite{ham05} analyzing the effective gap in weakly bound 
neutron levels in spherical and deformed nuclei, finding that the presence of 
a $s_{1/2}$ component in the wavefunction of these levels strongly reduces the 
effective gap. Other gap components with $l \ne 0$ show 
a less decreasing tendency near the drip line probably because of the 
centrifugal barrier which keeps the wave functions localised. This could mask
the behaviour of the state dependent gaps in real nuclei near the 
neutron drip line.

As mentioned already, another scenario where the quenching of the neutron 
gap appears is near the neutron drip line in the 
inner crust of neutron stars. This region is a crystal of nuclei
embeded in a gas of free neutrons and electrons. The inner crust was 
described by 
Negele and Vautherin \cite{neg73} at HF level by means of the energy density
functional method together with a spherical WS approach to deal 
with the crystal structure in an approximated way. 
The WS cell is electrically neutral and the ground state of the system
of neutrons, protons and electrons is reached when they are in 
$\beta$-equilibrium. 
We have performed a similar calculation but at TF 
level and using the BCP energy density functional \cite{bal08}. 
This functional consists of a bulk part provided by a microscopic 
calculation 
%\cite{bal04}
complemented by a phenomenological surface term. This functional with only 
four adjustable
parameters reproduces nuclear binding energies and charge radii of finite nuclei 
with the same quality as obtained with the most performant effective forces. 
The ground state energy per baryon obtained in this way for average 
densities in the 
WS cells ranging from $2.79 \times 10^{-4}$ to $7.89 \times 10^{-2}$ fm$^{-3}$ 
are displayed in the left panel of Fig. 2 (black dots) in comparison with the 
Negele-Vautherin 
results 
(red diamonds) finding an excellent agreement between both calculations. 
Although 
pairing correlations in the inner crust of neutron stars are mainly 
driven by the 
free neutron gas, they have, however, a noticeable influence on the 
composition and pairing
properties of the nuclear cluster inside the WS cell \cite{bal07}. 
On top of this TF calculation
in the inner crust, we have also performed a TF pairing calculation in 
the studied WS cells. To this end, we have used the Gogny D1S force 
renormalized by a factor
0.85 to take into account that in the BCP functional the effective 
mass equals the 
physical one. A specially relevant quantity in this context is the radial 
dependence of the pairing gap $\Delta({\bf R})$ which is obtained from 
the WT of the 
average quantal gap $\Delta_{av}=Tr[{\hat \Delta}{\hat 
\kappa}]/Tr[{\hat \kappa}]$ \cite{pill10}
which reads:
\begin{equation}
\Delta({\bf R}) = \frac{1}{\kappa({\bf R})} \int \frac{d {\bf p}}{(2\pi
\hbar)^3} \Delta({\bf R},{\bf p}) \kappa({\bf R},{\bf p})
\label{Delav0}
\end{equation}
where
%\begin{equation}
$\kappa({\bf R},{\bf p})=
 \int dE g^{TF}(E) \kappa(E) f_E({\bf R},{\bf p})$
and
 $\Delta({\bf R},{\bf p}) = -\int \frac{d {\bf p'}}{(2\pi \hbar)^3}
v({\bf p} - {\bf p'})
\kappa({\bf R},{\bf p'})$,
%\label{eq21}
%\end{equation}

 We see in the right panel of Fig. 2 that $\Delta({\bf R})$ takes a 
constant value in the outer part 
of the WS cell which corresponds to the gap of the free neutron gas 
\cite{cha10}. The predicted behaviour  for $\Delta({\bf R})$ are in  
qualitative agreement with previous calculations \cite{san04,bal07,gra08,
cha10}. The nuclear cluster inside the WS cell disappears when the 
homogeneous phase is reached at an average density of about 0.08 fm$^{-3}$ 
($^{982}$Ge$_{32}$). In this case the gap in the cell is 
practically the gap obtained in pure neutron matter at the same 
density. The density of the free neutron gas diminishes in approaching to
drip configurations. In this situation the gap is strongly reduced 
not only in the gas but also inside of the nuclear cluster ($^{200}$Nb$_{41}$)
and it even may disappear completely in the nucleus when the drip line 
is reached (see right panel of Fig. 1). 
%Notably, we 
%also find that the gap becomes very small or may even vanish even inside 
%the nucleus approaching drip configurations ($^{200}$Nb$_{41}$)
%in agreement with the result 
%displayed in the right panel of Fig. 1. 
Therefore, locally  
the TF $\Delta({\bf R})$'s are qualitatively different from what LDA would 
predict.

\section*{Conclusions and Outlook}
We have presented a TF theory for pairing in finite Fermi
systems for weak coupling situations where $\Delta/\varepsilon_F << 1$.
This TF theory differs from the
usual LDA. This essentially stems from the fact
that we approximate the gap equation in configuration space and, thus,
keep the size dependence of the matrix elements of the pairing force. This is
not the case in LDA where the matrix elements of the force are always
evaluated in plane wave basis.
This semiclassical approach to pairing is only based on the usual validity
criterion of Thomas-Fermi theory, namely that the Fermi wave length is smaller
than the oscillator length. At no point the LDA
condition that the coherence length must be
smaller than the oscillator
length enters the theory. Thus, the present TF approach yields for
all pairing quantities the same quality as TF theory does for
quantities in the normal fluid state.
An interesting feature of our study is that the average gap breaks down going
to the drip line. This unexpected result is confirmed by quantal calculations,
though strongly masked by shell fluctuations. For systems with large numbers 
of particles the fluctuations should die out and, thus, 
the semiclassical behaviour prevail. Indeed preliminary results in a slab 
configuration show good agreement between quantal and TF gaps around the drip 
region. We also investigated in slab geometry the inverse scenario where 
the external potential gets in the upper part suddenly strongly 
constricted rather than widened. 
Very preliminary 
results show that the gaps now become much enhanced where before they were 
suppressed. Putting such kind of slabs into a series could create a 
macroscopic system with strongly enhanced pairing properties. More studies of 
this kind are under way.

\section*{Acknowledgments}

We thank Michael Urban for useful discussions and comments.
This work has been partially supported by the IN2P3-CAICYT collaboration. 
One of us (X.V.) acknowledges grants FIS2008-01661 (Spain and FEDER),
2009SGR-1289 (Spain) and Consolider Ingenio Programme CSD2007-00042 for 
financial support.


\section*{References}

\begin{thebibliography}{99}
\bibitem{MS} W.D. Myers and W.J. \'{S}wi\c{a}tecki,
Ann. of Phys. {\bf 55}, 395 (1969); {\bf 84}, 186 (1974);
Nucl. Phys. {\bf A601}.
%\bibitem{MS1} W.D. Myers and W.J. \'{S}wi\c{a}tecki,
%Nucl. Phys. {\bf A601}, 147 (1996); Preprint LBL-36557 (1995).
\bibitem{far00} M. Farine, P. Schuck and X. Vi\~nas,
Phys. Rev. {\bf A62}, 013608 (2000); 
M. Durand, P. Schuck and J. Kunz, 
Nucl. Phys. {\bf 439}, 263 (1985).
\bibitem{Piek10} J. Duflo, Nucl. Phys. {\bf 576}, 29
(1994); J. Piekarewicz et al, 
%M. Centelles, X. Roca-Maza and X. Vi\~nas
 Eur. Phys. J. {\bf A46}, 379 (2010). 
\bibitem{vin03}
X. Vi\~nas, P. Schuck, M. Farine and M. Centelles,
Phys. Rev. {\bf C67}, 054307 (2003).
\bibitem{ham05}
I. Hamamoto, Phys. Rev. {\bf C71}, 037302 (2005).
\bibitem{san04}
N. Sandulescu, N. Van Giai and R.J. Liotta,
Phys. Rev. {\bf C69}, 045802 (2005).
\bibitem{bal07}
M. Baldo, U. Lombardo, E.E. Saperstein and S.V. Tolokonnikov,
Nucl. Phys. {\bf A750}, 409 (2005);
M. Baldo, E.E. Saperstein and S.V. Tolokonnikov,
Eur. Phys. J. {\bf A32}, 97 (2007).
\bibitem{gra08}
M. Grasso, E. Khan, J. Margueron and N. Van Giai,
Nucl. Phys. {\bf A807}, 1 (2008).
\bibitem{cha10}
N. Chamel, S. Goriely, J.M. Pearson and M. Onsi,
Phys. Rev. {\bf C81}, 045804 (2010).
\bibitem{kuch89}
H. Kucharek, P. Ring, P. Schuck,
R.Bengtsson and M. Girod,
Phys. Lett. {\bf B216}, 249 (1989).
\bibitem{RS} P. Ring and P. Schuck, {\it The Nuclear Many-Body Problem}, 
(Springer-Verlag, Berlin, 1980).
%\bibitem{gar99}
%E. Garrido, P. Sarriguren, E. Moya de Guerra and  P. Schuck,
%Phys. Rev. {\bf C60}, 064312 (1999);
\bibitem{D1S}
%J. Decharg\'e and D. Gogny,
%Phys. Rev. {\bf C21}, 1568 (1980);
J.-F. Berger, M. Giraud and D. Gogny,
Comp. Phys. Comm. {\bf 63}, 365 (1991).
%\bibitem{prak81}
%M. Prakash, S. Shlomo and V. A. Kolomietz,
%Nucl. Phys. {\bf 370}, 30 (1981).
\bibitem{cent98}
M. Centelles, X. Vi\~nas, M. Durand, P. Schuck and D. Von-Eiff,
Ann.of Phys. {\bf 266} 207 (1998).
\bibitem{soub00}
V.B. Soubbotin and X. Vi\~nas,
Nucl. Phys. {\bf A665}, 291  (2000).
\bibitem{soub03}
V.B. Soubbotin et al, 
%V.I. Tselyaev and X. Vi\~nas,
Phys. Rev. {\bf C67} 014324 (2003);
S. Krewald et al, 
%V.B. Soubbotin, V.I. Tselyaev and X. Vi\~nas,
Phys. Rev. {\bf C74} 064310 (2006).
\bibitem{vin11}
X. Vi\~nas, P. Schuck and M. Farine,
Int. J. Mod. Phys. {\bf E20} 399 (2011).
\bibitem{shl92}
S. Shlomo, Nucl. Phys. {\bf 539}, 17 (1992).
\bibitem{neg73}
J.W. Negele and D. Vautherin, Nucl. Phys. {\bf 207}, 298 (1973).
\bibitem{bal08}
M. Baldo, P. Schuck and X. Vi\~nas
Phys. Lett. {\bf B663}, 390 (2008).
%\bibitem{bal04}
%M. Baldo, C. Maieron, P. Schuck and X. Vi\~nas
%Nucl. Phys. {\bf 736}, 241 (2004).
\bibitem{pill10} 
%N. Pillet, N. Sandulescu and P. Schuck,
%Phys. Rev. {\bf C76} 024310 (2007);
N. Pillet, N. Sandulescu, P. Schuck and J.-F. Berger,
{\bf C81} 034307 (2010). 


\end{thebibliography}
\end{document}